\documentclass[english]{article}
\usepackage[T1]{fontenc}
\usepackage[latin9]{inputenc}
\usepackage{amsmath}
\usepackage{amssymb}
\usepackage{esint}

\makeatletter

\newcommand{\noun}[1]{\textsc{#1}}

\newcommand{\lyxaddress}[1]{
	\par {\raggedright #1
	\vspace{1.4em}
	\noindent\par}
}

\usepackage{babel}

\usepackage{babel}

\makeatother

\usepackage{babel}
\begin{document}
\title{Relevance of stochasticity for the emergence of quantization}
\date{A. M. Cetto, L. de la Peña, and A. Valdés-Hernández}
\maketitle

\lyxaddress{\vspace{-0.03\textheight}
Instituto de Física, Universidad Nacional Autónoma de México, Mexico
City}
\begin{abstract}
The theories of stochastic quantum mechanics and stochastic electrodynamics
bring to light important aspects of the quantum dynamics that are
concealed in the standard formalism. Here we take further previous
work regarding the connection between the two theories, to exhibit
the role of stochasticity and diffusion in the process leading from
the originally \emph{classical+}\emph{\noun{zpf}}\emph{ regime} to
the \emph{quantum regime}. Quantumlike phenomena present in other
instances in which a mechanical system is subject to an appropriate
oscillating background that introduces stochasticity, may point to
a more general appearance of quantization under such circumstances.
\end{abstract}

\section{Introduction}

In a recent paper \cite{arXiv20} we have discussed the connections
between stochastic quantum mechanics (\textsc{sqm}) and stochastic
electrodynamics (\textsc{sed}), two statistical theories that study
the dynamics of (otherwise classical) particles when embedded in a
stochastic environment. In essence, both theories are shown to provide
support in favor of a stochastic process underlying quantum mechanics.\footnote{There are of course a variety of theories containing a stochastic
element, aimed to explain or reproduce quantum mechanics. These may
widely differ from one another in their method, purpose or philosophy;
see, e. g., \cite{Fenyes 52}-\cite{Lindgren 2019}. It is not our
intention to review such theories, but to draw on the specific connections
between \textsc{sqm} and \textsc{sed} for a better understanding of
the role of stochasticity in the emergence of quantization.}

The results of \noun{sqm} and \noun{sed} suggest that for a more satisfactory
understanding of the mechanism of quantization it is essential to
consider that matter is in permanent interaction with a physical medium
that brings about a stochastic behavior of the system. While \noun{sqm}
does not specify the nature of such medium, \noun{sed} in particular
identifies it with the zero-point radiation field (\noun{zpf}). This
is an ubiquitous random electromagnetic radiation field with energy
per mode $\hbar\omega/2$, which accounts for the appearance of Planck's
constant and the wave element in quantum mechanics, as well as Born's
rule associated with it \cite{Dice,TEQ}. The \noun{zpf} has been
shown to play an essential role in producing quantum features such
as the so-called quantum indeterminism \cite{Indet2009}, entanglement
\cite{Entangle,Khrennikov14,TEQ} and others.

The present paper is devoted to a more in-depth discussion of the
transit from the originally classical+\noun{zpf} regime to the quantum
regime, and of the crucial{\small{} }role played by diffusion in bringing
about such a qualitative change in the dynamics. The discussion suggests
investigating other instances in which the permanent interaction of
a particle with a vibrating field introduces a stochastic element
into the dynamics, to look for possible signs pointing to the generality
of quantization under such circumstances.

The structure of the paper is as follows. To set the framework, in
section \ref{sed} we succinctly recall the \textsc{sed} (statistical)
treatment of a particle subject to an external potential, leading
to a description in configuration space. In section \ref{emsm} we
briefly introduce the basic equations of \textsc{sqm} and, by linking
with the \noun{sed} description, we complete the equations of \textsc{sqm}
with the inclusion of the radiative corrections. We further show how
the \textsc{sed} and \textsc{sqm} equations connect with the Schrödinger
equation, and relate the quantum momentum operator with the local
average velocities that are central elements in the \textsc{sed} and
\textsc{sqm} approaches. This allows to stress the role of the \textsc{zpf}-induced
diffusion in driving the system to its non-classical behavior. In
Section \ref{c2q}, a clue to understand the mechanism leading to
a (statistical) description of the quantum regime in terms of operators
and state vectors is put forward. The paper ends with a discussion
on the relevance of the wave element associated with the source of
stochasticity, which is also present in the walking-droplet systems
that exhibit a quantumlike behavior \cite{Coud20XX,Bush2015}. 

\section{The underlying equations of SED\label{sed}}

\noun{Sed} provides a statistical description of the dynamics of a
charged particle (typically an atomic electron) in interaction with
the \textsc{zpf}, subject to an external potential and possibly to
an external electromagnetic field. The conventional starting point
of the theory is the Langevin equation, also known in \textsc{sed}
as Braffort-Marshall equation, which is the nonrelativistic dipole
approximation of the (stochastic) Abraham-Lorentz equation (for simplicity
we consider the case in which there is no external radiation field)
\begin{equation}
m\boldsymbol{\ddot{x}}=\boldsymbol{f}(\boldsymbol{x})+m\tau\boldsymbol{\dddot{x}}+e\boldsymbol{E},\label{8}
\end{equation}
where $m\tau\boldsymbol{\dddot{x}}$ represents the radiation reaction
force with $\tau=2e^{2}/3mc^{3}$ ($\sim$ 10$^{-23}$ s for the electron),
$\boldsymbol{f}=-\boldsymbol{\nabla}V(\boldsymbol{x})$, and $\boldsymbol{E}$
represents the electric component of the (random) \textsc{zpf}. The
latter is usually ---but not necessarily--- taken in the dipole
approximation, $\boldsymbol{E}(t)$.\footnote{This approximation is justified a posteriori \cite{TEQ}, as it turns
out that the wavelength of the relevant modes of the \textsc{zpf},
i. e., those with which the particle interacts resonantly in the quantum
regime (see below), is indeed much larger than the displacements of
the electron around its mean position.} 

Since the \textsc{zpf} is an intrinsic component of the \textsc{sed}
system, the canonical momentum is defind simply as $\boldsymbol{p}=m\boldsymbol{\dot{x}}$,
whence the equation of motion becomes
\begin{equation}
\boldsymbol{\dot{p}}=\boldsymbol{f}+m\tau\boldsymbol{\dddot{x}}+e\boldsymbol{E}(t).\label{13}
\end{equation}

In the absence of the \textsc{zpf} we would have a purely \emph{classical}
electrodynamic problem. The presence of the term $e\boldsymbol{E}$
renders the problem stochastic and therefore amenable to a statistical
treatment only. A standard procedure (see e. g. \cite{Ris84}, \cite{TEQ}
Ch. 4) leads to a generalized Fokker-Planck equation (\textsc{gfpe})
for the phase-space probability distribution $Q(\boldsymbol{x},\boldsymbol{p},t)$
with a memory term, 
\begin{equation}
\left(\hat{L}_{c}+e^{2}\hat{L}_{r}\right)Q(\boldsymbol{x},\boldsymbol{p},t)=0,\label{16}
\end{equation}
where 
\begin{equation}
\hat{L}_{c}=\frac{\partial}{\partial t}+\frac{1}{m}\boldsymbol{\nabla}\cdot\boldsymbol{p}+\boldsymbol{\nabla}_{p}\cdot\boldsymbol{f}\label{17}
\end{equation}
stands for the classical Liouvillian and 
\begin{equation}
\hat{L}_{r}=\boldsymbol{\nabla}_{p}\cdot\left(\frac{2}{3c^{3}}\boldsymbol{\dddot{x}}-\mathcal{\hat{\boldsymbol{D}}}\right)\label{18}
\end{equation}
for the radiative and diffusive terms. The latter contains the integro-differential
operator, written here to lowest order in $e^{2}$, 
\begin{equation}
\mathcal{\hat{\boldsymbol{D}}}=\intop_{-\infty}^{t}dt^{\prime}\varphi(t-t^{\prime})\nabla_{p^{\prime}},\quad\varphi(t)=\frac{2\hbar}{3\pi c^{3}}\int_{0}^{\infty}d\omega\,\omega^{3}\cos\omega t,\label{19}
\end{equation}
where $\boldsymbol{p}^{\prime}=\boldsymbol{p}(t^{\prime})$ is the
value of the momentum at $t^{\prime}<t$, such that it evolves towards
$\boldsymbol{p}=\boldsymbol{p}(t)$, and $\varphi(t)$ stands for
the \textsc{zpf} covariance.

Equation (\ref{16}) describes the evolution of the phase-space probability
density at all times. It is virtually impossible to construct its
general solution, yet an approximate method has been developed that
leads to a good (approximate) description for asymptotic times, when
the average effects of the radiative and diffusive terms compensate
each other, and they become then small radiative corrections (see
\cite{TEQ} and references therein). As shown in detail in \cite{PeCeVa14}
(or \cite{TEQ} Ch. 4), it is found that the corresponding (asymptotic)
regime can be identified with the \emph{quantum regime} (see Section
\ref{transit}). 

In order to connect with \textsc{sqm}, we first reduce the \textsc{sed}
description to the configuration space by integrating over the momentum.
The \emph{local} mean value of a dynamical variable $G(\boldsymbol{x},\boldsymbol{p})$
is given by 
\begin{equation}
\langle G\rangle_{x}\equiv\frac{1}{\rho}\int d\boldsymbol{p}\,G(\boldsymbol{x},\boldsymbol{p})Q(\boldsymbol{x},\boldsymbol{p},t),\label{local}
\end{equation}
where $\rho=\rho(\boldsymbol{x},t)=\int d\boldsymbol{p}\,Q(\boldsymbol{x},\boldsymbol{p},t)$
stands for the probability density of particles in $\boldsymbol{x}$-space.
The equation of evolution for $\langle G\rangle_{x}$ is obtained
by left-multiplying the \textsc{gfpe} by $G$ before integrating over
$\boldsymbol{p}$. For $G=1$, integration of (\ref{16}) gives the
continuity equation 
\begin{equation}
\frac{\partial\rho}{\partial t}+\boldsymbol{\nabla}\cdot\left(\rho\boldsymbol{v}\right)=0,\label{20}
\end{equation}
with $\boldsymbol{v}=\boldsymbol{v}(\boldsymbol{x},t)$ the flux (or
current) velocity, 
\begin{equation}
\boldsymbol{v}=\langle\boldsymbol{\dot{x}}\rangle_{x}=\frac{1}{m}\langle\boldsymbol{p}\rangle_{x}.\label{21}
\end{equation}
For $G=\boldsymbol{p}$ one gets, summing over repeated indices
\begin{equation}
\frac{\partial}{\partial t}m\boldsymbol{v}\rho+m^{2}\partial_{j}\left\langle \dot{x}_{j}\boldsymbol{\dot{x}}\right\rangle _{x}\rho-\left\langle \boldsymbol{f}\right\rangle _{x}\rho=e^{2}\left(\frac{2}{3c^{3}}\left\langle \boldsymbol{\dddot{x}}\right\rangle _{x}-\langle\mathcal{\hat{\boldsymbol{D}}}\rangle_{x}\right)\rho\equiv\boldsymbol{R}_{x}.\label{22}
\end{equation}

\section{Connecting with SQM\label{emsm}\textsc{\label{sqm}}}

\noun{Sqm} describes the dynamics of a particle undergoing a stochastic
motion, without further inquiry about the source of the stochasticity.
The general equations of motion correspond to the time-inversion invariant
and non-invariant stochastic extension, respectively, of Newton's
equation of motion, namely \cite{arXiv20,Dice} 
\begin{equation}
m\left(\mathcal{\hat{D}}_{c}\boldsymbol{v}-\lambda\mathcal{\hat{D}}_{s}\boldsymbol{u}\right)=\boldsymbol{f}_{+},\quad m\left(\mathcal{\hat{D}}_{c}\boldsymbol{u}+\mathcal{\hat{D}}_{s}\boldsymbol{v}\right)=\boldsymbol{f}_{-},\label{A10}
\end{equation}
where $\lambda$ is a real parameter that can be taken as $\lambda^{2}=1$,
\begin{equation}
\mathcal{\hat{D}}_{c}=\frac{\partial}{\partial t}+\boldsymbol{v}\cdot\boldsymbol{\nabla},\quad\mathcal{\hat{D}}_{s}=\boldsymbol{u}\cdot\boldsymbol{\nabla}+D\boldsymbol{\nabla}^{2}\label{A11}
\end{equation}
are the so-called systematic and stochastic derivatives, respectively,
$\boldsymbol{v}$ is the flux velocity, and $\boldsymbol{u}$ is the
diffusive (or stochastic) velocity 
\begin{equation}
\boldsymbol{u}(\boldsymbol{x},t)=D\frac{\boldsymbol{\nabla}\rho}{\rho},\label{A12}
\end{equation}
with $\rho(\boldsymbol{x},t)$ the probability density of particles
and $D$ the diffusion constant. The forces $\boldsymbol{f}_{-}$
and $\boldsymbol{f}_{+}$ in Eqs. (\ref{A10}) do and do not change
sign, respectively, under a time inversion. The Newtonian limit (the
classical Hamiltonian description) corresponds to $D=0$, hence $\boldsymbol{u}=\boldsymbol{0}$,
which means no diffusion at all.

Equations (\ref{A10}) can be further combined into the single, compact
equation 
\begin{equation}
\mathcal{\hat{D}}_{\kappa}\boldsymbol{p}_{\kappa}=\boldsymbol{f}_{\kappa},\label{A22}
\end{equation}
with
\begin{equation}
\boldsymbol{p}_{\kappa}=m\boldsymbol{w},\boldsymbol{\quad w}=\boldsymbol{v}-\sqrt{-\lambda}\boldsymbol{u},\quad\boldsymbol{f}_{\kappa}=\boldsymbol{f}_{+}-\sqrt{-\lambda}\boldsymbol{f}_{-},\label{A22a}
\end{equation}
\begin{equation}
\mathcal{\hat{D}}_{\kappa}=\mathcal{\hat{D}}_{c}-\sqrt{-\lambda}\mathcal{\hat{D}}_{s}=\frac{\partial}{\partial t}+\frac{1}{m}\boldsymbol{p}_{\kappa}\cdot\boldsymbol{\nabla}-\sqrt{-\lambda}\,D\,\boldsymbol{\nabla}^{2}.\label{A22b}
\end{equation}

The sign of $\lambda$ serves to distinguish between the two basic
stochastic processes in the Markovian approximation (see e.g. \cite{2015Spec}):
$\lambda=-1$ corresponds to the classical case (Brownian motion),
and $\lambda=1$ to the quantum case. Therefore, in what follows we
take $\lambda=1$. This means, in particular, that (see Eq. (\ref{A22a}))
\begin{equation}
\boldsymbol{w}=\boldsymbol{v}-i\boldsymbol{u}.\label{w}
\end{equation}
We shall come back to this important equation below.

\subsection{From \textsc{sed} to \textsc{sqm}}

The link between the two theories is established by introducing the
\noun{sqm} expressions for the velocity $\boldsymbol{u},$ Eq. (\ref{A12}),
and the coefficient \cite{Dice} 
\begin{equation}
D=\frac{\hbar}{2m},\label{A13}
\end{equation}
into the \textsc{sed} equation (\ref{22}), and combining this with
(\ref{20}) to obtain
\begin{equation}
mv_{i}\left(\frac{2m}{\hbar}\boldsymbol{u}\cdot\boldsymbol{v}+\boldsymbol{\nabla}\cdot\boldsymbol{v}\right)+m\left(\frac{2m}{\hbar}\boldsymbol{u}\cdot\left\langle \boldsymbol{\dot{x}}\dot{x}_{i}\right\rangle _{x}+\boldsymbol{\nabla}\cdot\left\langle \boldsymbol{\dot{x}}\dot{x}_{i}\right\rangle _{x}\right)=f_{i}+R_{xi}.\label{32-1}
\end{equation}
for every Cartesian component $\emph{i}$. This suggests introducing
the tensor $T_{ij}$, given by the (local) correlation between the
components of the vector $\boldsymbol{\dot{x}}$, 
\begin{equation}
T_{ij}=-\frac{2m}{\hbar}\left(\left\langle \dot{x}_{i}\dot{x}_{j}\right\rangle _{x}-v_{i}v_{j}\right)=-\frac{2m}{\hbar}\left(\left\langle \dot{x}_{i}\dot{x}_{j}\right\rangle _{x}-\left\langle \dot{x}_{i}\right\rangle _{x}\left\langle \dot{x}_{j}\right\rangle _{x}\right).\label{34}
\end{equation}
Equation (\ref{32-1}) takes then the simpler form (summing over repeated
indices) 
\begin{equation}
m\left(\frac{\partial v_{i}}{\partial t}+v_{j}\partial_{j}v_{i}-T_{ij}u_{j}-\frac{\hbar}{2m}\partial_{j}T_{ij}\right)=f_{i}+R_{xi}.\label{36}
\end{equation}
Notice that when $R_{xi}$ is neglected, this equation is equivalent
to the first equation in (\ref{A10}) with $\lambda=1$, $\boldsymbol{f}_{+}=\boldsymbol{f}$,
and $T_{ij}$ given by 
\begin{equation}
T_{ij}=\partial_{j}u_{i}=D\partial_{j}\partial_{i}\ln\rho=T_{ji},\label{38}
\end{equation}
hence $T_{ij}$ plays the role of a stress rate tensor associated
with the local mean changes of $\boldsymbol{u}.$ By combining this
with Eqs. (\ref{21}), (\ref{A12}), (\ref{A13}) and (\ref{34})
we get
\begin{equation}
\langle p_{i}p_{j}\rangle_{x}-\langle p_{i}\rangle_{x}\langle p_{j}\rangle_{x}=\frac{\hbar^{2}}{4}\partial_{i}\partial_{j}\ln\rho=m^{2}DT_{ij},\label{A33}
\end{equation}
which points to the significance of $T_{ij}$, in particular of its
trace $\sum_{i}T_{ii}=(m^{2}D)^{-1}\sigma_{p}^{2}(\boldsymbol{x})$.
Notice that the local dispersion of the momentum, $\sigma_{p}^{2}(\boldsymbol{x})$,
is determined by the (divergence of the) diffusive velocity alone.

Further, the \textsc{sed} approach has provided us the means to arrive
at the \emph{complete} dynamical equation of \textsc{sqm} with the
radiative terms included,
\begin{equation}
\mathcal{\hat{D}}_{\kappa}\boldsymbol{p}_{\kappa}=\boldsymbol{f}_{\kappa}\boldsymbol{}+\boldsymbol{R}_{x}.\label{39a}
\end{equation}

\subsection{Final step: the quantum description}

\label{transit} Connecting with the Schrödinger equation is now essentially
an algebraic exercise, which is accomplished by introducing the complex
function $\psi$ such that 
\begin{equation}
\boldsymbol{p}_{\kappa}=m\boldsymbol{(v}-i\boldsymbol{u})=\hbar\,(\mathrm{Im}\frac{\boldsymbol{\nabla}\psi}{\psi}-i\,\mathrm{Re}\frac{\boldsymbol{\nabla}\psi}{\psi})=-i\hbar\,\frac{\boldsymbol{\nabla}\psi}{\psi}\label{40}
\end{equation}
into Eq. (\ref{39a}) in the radiationless regime (i.e., without the
term $\boldsymbol{R}_{x}$) and integrating once, to obtain\footnote{More detailed derivations are presented in \cite{TEQ} and references
therein. Further, the radiative terms neglected here have been explored
to the lowest order of approximation and shown to reproduce the predictions
of nonrelativistic quantum electrodynamics \cite{CePe 91}-\cite{CePeVa 13}.} 
\begin{equation}
\frac{1}{2m}\left(-i\hbar\boldsymbol{\nabla}\right)^{2}\psi+V\psi=i\hbar\frac{\partial\psi}{\partial t}.\label{A36}
\end{equation}

Note that according to Eq. (\ref{40}), the momentum operator $\boldsymbol{\hat{p}}=-i\hbar\boldsymbol{\nabla}$
is directly related with the (complex) velocity $\boldsymbol{w}$
of \textsc{sqm}, 
\begin{equation}
\boldsymbol{\hat{p}}\psi=-i\hbar\,\boldsymbol{\nabla}\psi=m(\boldsymbol{v}-i\boldsymbol{u})\psi=m\boldsymbol{w}\,\psi,\label{31A}
\end{equation}
which shows that $\boldsymbol{w}$ is the relevant velocity, or rather,
that the two velocity components $\boldsymbol{v}$ and $\boldsymbol{u}$
play an equally important role in the dynamics. The fact that they
both contribute to the average energy can be utilized to derive the
time-independent Schrödinger equation from a variational principle.
Indeed, from Eq. (\ref{31A}) the average kinetic energy can be written
as 
\begin{eqnarray}
\langle T\rangle & = & \frac{1}{2}m\intop(\boldsymbol{v}^{2}+\boldsymbol{u}^{2})\rho\,d\boldsymbol{x}=\frac{1}{2}m\intop(\boldsymbol{w}\psi)\cdot(\boldsymbol{w}^{*}\psi^{*})\,d\boldsymbol{x}\nonumber \\
 & = & \frac{\hbar^{2}}{2m}\intop(\boldsymbol{\nabla}\psi)\cdot(\boldsymbol{\nabla}\psi^{*})\,d\boldsymbol{x}.
\end{eqnarray}
 The total average energy is then
\begin{equation}
E=\intop\left[\frac{\hbar^{2}}{2m}(\boldsymbol{\nabla}\psi^{*})\cdot\left(\boldsymbol{\nabla}\psi\right)+V\psi^{*}\psi\right]d\boldsymbol{x},\label{A6n}
\end{equation}
and by varying both $\psi^{*}$ and $\psi$, subject to the constraint
imposed by the normalization condition, i. e., $\delta N=\intop\left(\psi^{*}\delta\psi+\psi\delta\psi^{*}\right)d\boldsymbol{x}=0,$
one obtains after an integration by parts, assuming a bounded system,
\begin{equation}
\delta E=\intop\left\{ \left[-\frac{\hbar^{2}}{2m}\boldsymbol{\nabla}^{2}\psi^{*}+(V-\gamma)\psi^{*}\right]\delta\psi+\left[-\frac{\hbar^{2}}{2m}\boldsymbol{\nabla}^{2}\psi+(V-\gamma)\psi\right]\delta\psi^{*}\right\} d\boldsymbol{x},\label{A8n}
\end{equation}
with $\gamma$ a free real parameter. Since $\delta N=0,$ one may
add a term $(\gamma-E)\left(\psi^{*}\delta\psi+\psi\delta\psi^{*}\right)$
to the integrand of (\ref{A8n}). The condition $\delta E=0$ implies
thus the (stationary) Schrödinger equation 
\begin{equation}
-\frac{\hbar^{2}}{2m}\boldsymbol{\nabla}^{2}\psi+V\psi=E\psi\label{A10n}
\end{equation}
and its complex conjugate, with $E$ given by Eq. (\ref{A6n}).

\section{On the emergence of the quantum behavior \label{c2q}}

The above results highlight the importance of diffusion in eliciting
a non-classical behaviour of the system. In what follows we delve
into the physical mechanism by which the \textsc{zpf}-induced diffusion
leads to the quantum regime. 

\subsection{The equilibrium condition. Effect of diffusion on the dynamics}

Looking back at the \textsc{gfpe}\noun{, }Eq. (\ref{16}), one can
see that what makes the system behave nonclassically are the two terms
contained in $\hat{L}_{r}$, which means they deserve closer inspection.
For this purpose we multiply (\ref{16}) by any constant of the motion
$G(\boldsymbol{x},\boldsymbol{p})=\xi$ and integrate over $\boldsymbol{p}$.
The terms associated with the classical Liouvillian, $\hat{L}_{c}\xi$,
cancel out, and those associated with $\hat{L}_{r}\xi$ must balance
each other on average by virtue of the equilibrium condition $\langle d\xi/dt\rangle=0$,
i. e., (here $\boldsymbol{g}(\boldsymbol{x},\boldsymbol{p})=\boldsymbol{\nabla}_{p}\xi(\boldsymbol{x},\boldsymbol{p})$)
\begin{equation}
-\left\langle \dddot{\boldsymbol{x}}\cdot\boldsymbol{g}\right\rangle =\frac{\hbar}{\pi}\int_{0}^{\infty}d\omega\,\omega^{3}\int_{-\infty}^{t}dt\cos\omega(t-t^{\prime})\left\langle \boldsymbol{\nabla}_{p^{\prime}}\cdot\boldsymbol{g}\right\rangle ,\label{10-1}
\end{equation}
where, as said before, $\boldsymbol{p}^{\prime}=\boldsymbol{p}(t^{\prime})$
is the value of the momentum at $t^{\prime}<t$, such that it evolves
towards $\boldsymbol{p}=\boldsymbol{p}(t)$, This constitutes a strong
condition on the dynamics; it implies that only those solutions of
the radiationless (zero-order) part of the \textsc{gfpe} that satisfy
this condition, are valid solutions in the equilibrium regime. 

Although Eq. (\ref{10-1}) holds only under stationarity, each side
of it can be analyzed separately at all times. The l.h.s. term is
due to radiation reaction, and therefore represents the dissipative
part. Take for instance the case in which $\xi$ represents the energy.
Initially (at $t=-\infty$, when particles and \textsc{zpf} start
to interact) the dissipative term obviously dominates over the diffusive
one. Were it not for the r.h.s. term, the particles would eventually
exhaust their energy and come to a complete standstill. As time progresses,
however, the diffusion of the momentum increases thanks to the action
of the \textsc{zpf}, until it reaches a point where the r.h.s. term
does not depend on the time variable; this is the Markovian limit,
well described by the equations of \textsc{sqm} with the second-order
derivative term containing a constant diffusion coefficient $D$.

The factor $\boldsymbol{\nabla}_{p^{\prime}}\cdot\boldsymbol{g}$
is at the core of the mechanism of evolution towards the balance regime;
it signals the effects of the diffusion of $\boldsymbol{p}$ due to
the \textsc{zpf}. Let us consider the upper time limit of the integral,
$t,$ close to the asymptotic time. For small values of $t^{\prime}$
(close to the lower limit), the behavior of $\boldsymbol{p}(t^{\prime})$
is largely classical (not diffusive) and differs markedly from that
of $\boldsymbol{p}(t)$; yet this difference is blurred as time progresses.
In classical mechanics (in the absence of diffusion), the quantity
$\boldsymbol{\nabla}_{p^{\prime}}\cdot\boldsymbol{g}$ can be expressed
in terms of the Poisson bracket 

\begin{equation}
\frac{\partial g_{i}(t)}{\partial p_{i}(t^{\prime})}=\left\{ x_{i}(t^{\prime}),g_{i}(t)\right\} .\label{52}
\end{equation}
This represents an abridged description of the classical evolution,
which is purely deterministic, as opposed to that described by the
\noun{gfpe}, which is \emph{statistically} deterministic, meaning
that although the motion of individual particles follows deterministic
rules, the evolution of the ensemble is defined only in a statistical
sense. The r.h.s. of Eq. (\ref{10-1}) ---and with it the entire
equation--- ceases to follow classical Hamiltonian laws as soon as
the diffusion enters into force. The new dynamics should reflect as
a fundamental property the role played by diffusion (which in usual
quantum mechanics is considered under the notion of indeterminism).
In the asymptotic limit, the expression $\boldsymbol{\nabla}_{p^{\prime}}\cdot\boldsymbol{g}$,
and the corresponding symplectic structure represented in the classical
case by the Poisson bracket, must capture this fundamental change
in the dynamics. This, in essence, is what justifies in \noun{sed}
the (otherwise pragmatic) transition from the Poisson bracket to the
corresponding commutator, which serves to express in a language proper
of a statistical treatment the meaning of the quantity $\boldsymbol{\nabla}_{p^{\prime}}\cdot\boldsymbol{g}$:
\begin{equation}
\frac{\partial g_{i}(t)}{\partial p_{j}(t^{\prime})}\rightarrow\beta\left[\hat{x}_{i}(t^{\prime}),\hat{g}_{j}(t)\right],\label{54}
\end{equation}
with the operators acting on the state function $\psi$ that represents
the ensemble under consideration. The value of the parameter $\beta$
is to be determined by the balance condition (\ref{10-1}). For this
purpose we apply consistently in Eq. (\ref{10-1}) the substitution
rule (\ref{54}), which for $t'=t$ and $\hat{g}_{j}=\hat{p}_{j}$
means 
\begin{equation}
\left\{ x_{i},p_{j}\right\} =\delta_{ij}\rightarrow\beta\left[\hat{x}_{i},\hat{p}_{j}\right]=\delta_{ij},\label{58}
\end{equation}
and take an average over the ensemble of systems in the ground state,
viz. the (sole) state in equilibrium with the \textsc{zpf}. For the
sake of clarity we restrict here the calculations to one dimension.
With the matrix elements $x_{kn}(t)=x_{kn}\exp(i\omega_{kn}t)$, $g_{kn}=-\beta(\xi_{k}-\xi_{n})x_{kn}$,
and taking into account that $\xi_{kn}=\xi_{n}\delta_{kn}$ for any
constant $\xi$, we obtain for the l.h.s. of (\ref{10-1}), 
\begin{equation}
-\left\langle \dddot{\boldsymbol{x}}\cdot\boldsymbol{g}\right\rangle _{0}=-i\beta\sum_{k}(\xi_{k}-\xi_{0})\omega_{k0}^{3}\left\vert x_{k0}\right\vert ^{2},\label{60}
\end{equation}
which does not explicitly contain Planck's constant. For the r.h.s.
we get
\begin{equation}
\frac{\hbar}{\pi}\int_{0}^{\infty}d\omega\,\omega^{3}\int_{-\infty}^{t}dt\cos\omega(t-t^{\prime})\left\langle \boldsymbol{\nabla}_{p^{\prime}}\cdot\boldsymbol{g}\right\rangle _{0}=\hbar\beta^{2}\sum_{k}(\xi_{k}-\xi_{0})\omega_{k0}^{3}\left\vert x_{k0}\right\vert ^{2}.\label{62}
\end{equation}
The balance condition reads therefore
\begin{equation}
-i\beta\sum_{k}(\xi_{k}-\xi_{0})\omega_{k0}^{3}\left\vert x_{k0}\right\vert ^{2}=\hbar\beta^{2}\sum_{k}(\xi_{k}-\xi_{0})\omega_{k0}^{3}\left\vert x_{k0}\right\vert ^{2},\label{63}
\end{equation}
whence $\beta=-i/\hbar$, and consequently,
\begin{equation}
\left[\hat{x}_{i},\hat{p}_{j}\right]=i\hbar\delta_{ij}.\label{66}
\end{equation}
This result encapsulates in a most remarkable form the profound effect
of the \textsc{zpf} on the dynamics. \textsc{sed} endows thus the
formal rule $\left\{ f,g\right\} \rightarrow\frac{1}{i\hbar}\left[\hat{f},\hat{g}\right]$
with a deep physical sense: $\hbar$ is the hallmark of the \textsc{zpf}.
The new dynamics ---an extension of Hamiltonian dynamics that embodies
the effects of fluctuations--- becomes expressed in terms of operators,
and refers no more to trajectories of particles moving in ordinary
space, but to a statistical ensemble of them in a given state.\footnote{This is not a unique situation in theoretical physics; there are several
(although related) examples in which stochasticity brings about a
qualitative change in the dynamics. The title of section 2.4 of the
forerunner paper by Chandrasekhar \cite{Chandra} reads ``The Fokker-Planck
Equation. The Generalization of Liouville's Theorem.'' The generalization
at issue is just an extension of the Hamiltonian dynamics of Liouville's
theorem, so as to embed fluctuations and dissipation into the scheme.
A more recent example, closer to our case, is the discovery by Nelson
\cite{Nelson 1966} of the need of two velocities for an appropriate
description of the dynamics of a stochastic system, just as discussed
above. However. Nelson continued to call his theory Newtonian. } The physical description leaps from ordinary space into an abstract
Hilbert space; from a transparent visualization into a formal representation.
Here the descriptions of Schrödinger and Heisenberg converge; the
ensuing description is statistical (although neither of these authors
knew it at their time), with wave functions living in the Hilbert
space on which the operators act.\footnote{In a separate work \cite{PeCeVa09} it been shown that under conditions
of ergodicity, the dynamical variables describing the statistical
properties of an ensemble in a given (pure) state are expressed by
the corresponding quantum operators; this represents a complementary
derivation of quantum mechanics \emph{à la} Heisenberg.}

\subsection{The undulatory element}

The wavelike nature of the \textsc{zpf} and the strong self-correlation
of its modes, as manifest in the field covariance (\ref{19}), are
features that distinguish the quantum case from Brownian motion. As
discussed above, initially the dynamics of the \textsc{sed} system
is irreversible, until a balance is reached between the average effects
of diffusion and dissipation; at that point the dynamics has become
reversible and can be described in terms of stationary solutions.
This suggests that it should be possible to imagine other instances
in which a (material) system is acted on by a permanent oscillatory
background field that induces a stochastic response; does this always
imply that the system acquires wavelike properties and eventually
reaches a regime characterized by a quantumlike behavior? It remains
to investigate to what extent such a qualitative change in the dynamics
can be reproduced (or observed) in other instances where an otherwise
`classical' system is subject to a similar combination of the undulatory
and the stochastic element, giving rise to quantization. 

This question gains relevance in light of the recent series of remarkable
experimental, theoretical and numerical work carried out in hydrodynamics,
with droplets bouncing on the surface of a vibrating fluid and describing
trajectories guided by their accompanying surface waves \cite{Coud20XX},
which thus constitute a sort of hydrodynamic de Broglie waves. A number
of phenomena have been observed with one or more walking droplets,
which show clear signs of interference effects, nonlocal interaction
between droplets, and quantization of orbits (\cite{Coud20XX}, \cite{Bush2015},
\cite{nachbin}, \cite{Turton 18} and references therein). The detailed
equations that govern such systems have all imaginable complexities;
yet in the end, the smoothened-out, averaged trajectories of the droplets
are observed to form regular patterns that strongly suggest an analogy
with quantum mechanics. It should be possible, therefore, to find
some way of transiting from the detailed description of the coupled
(droplet$+$surface) system to an approximate, averaged description
of the droplet motion that serves to determine the extent of the quantum
analogy and the conditions under which it manifests itself.

In this regard, it is pertinent to remark that Eq. (\ref{63}) is
satisfied frequency by frequency; it expresses a \emph{detailed balance}
\cite{TEQ}. The balance condition can therefore be considered as
a kind of fluctuation-dissipation theorem. Whether an equivalent balance
condition can be established in the hydrodynamic case is a matter
for further investigation. By pursuing this analogy, one should be
able to learn more about the process of quantization. In particular,
the fact that in the (macroscopic) hydrodynamic case the trajectories
can be visualized and recorded, should help in investigating the transient
phase, when the `quantum regime' has not yet taken over. In the context
of the quantum analogy, this would be equivalent to testing the \textsc{sed}
predictions by taking the system out of the quantum regime. 

Acknowledgments. We thank the reviewers for their helpful and valuable
comments. Partial support from DGAPA-UNAM through project PAPIIT IN113720
is acknowledged.

All authors have made substantial contributions to this paper.

\end{document}